\documentclass[reprint,superscriptaddress,amsmath,amssymb,aps,prb,longbibliography]{revtex4-2}
\usepackage[utf8]{inputenc}
\usepackage{graphicx}
\usepackage{dcolumn}
\usepackage{bm}
\usepackage[colorlinks,citecolor=blue,linkcolor=blue,urlcolor=blue]{hyperref}
\usepackage[normalem]{ulem}

\begin{document}

\title{Oriented Triplet $p$-Wave Pairing from  Fermi surface Anisotropy and Nonlocal Attraction}

 \author{Shuning Tan}
\thanks{These authors contributed equally.}
\affiliation{Key Laboratory for Microstructural Material Physics of Hebei Province, School of Science, Yanshan University, Qinhuangdao 066004, China} 

 \author{Ji Liu}
 \thanks{These authors contributed equally.}
 \affiliation{School of Science, Harbin Institute of Technology, Shenzhen, 518055, China}
 \affiliation{Shenzhen Key Laboratory of Advanced Functional Carbon Materials Research and Comprehensive Application, Shenzhen 518055, China.}

\author{Minghuan Zeng}
\affiliation{College of Physics, Chongqing University, Chongqing 401331, China}

\author{Tao Ying}
\affiliation{School of Physics, Harbin Institute of Technology, Harbin 150001, China}

\author{Zhangkai Cao}
\email{caozhangkai@mail.bnu.edu.cn}
\affiliation{Eastern Institute for Advanced Study, Eastern Institute of Technology, Ningbo, Zhejiang 315200, China}
\affiliation{School of Physical Sciences, University of Science and Technology of China, Hefei, 230026, China}

\author{Ho-Kin Tang}
\email{denghaojian@hit.edu.cn}
\affiliation{School of Science, Harbin Institute of Technology, Shenzhen, 518055, China}
\affiliation{Shenzhen Key Laboratory of Advanced Functional Carbon Materials Research and Comprehensive Application, Shenzhen 518055, China.}

\date{\today}

\begin{abstract}
Using constrained-path quantum Monte Carlo, we map the ground-state phase diagram versus the nearest-neighbor (NN) attraction $V$ and spin-dependent hopping anisotropy $\alpha$ for the two-dimensional attractive $t$--$U$--$V$ Hubbard model at filling $n\simeq0.85$.
We identify an onsite $s$-wave superfluid, a Cooper pair Bose metal with an uncondensed Bose surface, and an oriented equal-spin triplet $p$-wave pairing phase.
The NN attraction activates the odd-parity channel, while hopping anisotropy suppresses the competing $s$-wave coherence and selects a $p_x/p_y$ polar axis, 
and thus lowers the critical $|V_c|$ for the onset of triplet-dominant $p$-wave pairing.
A channel-resolved Landau analysis provides a  criterion for the
Landau $p$-wave scale $V_c^{\mathrm L}(\alpha)$, consistent with the observed anisotropy dependence of $|V_c|$.
Our results establish how NN interaction and  Fermi surface anisotropy cooperate to generate the oriented triplet $p$-wave pairing, 
and suggest that cold-atom and altermagnetic platforms could potentially realize this mechanism.
\end{abstract}

\maketitle
Spin-triplet $p$-wave pairing has long been pursued as a route to odd-parity superconductivity and topological superfluids hosting Majorana modes~\cite{gu2025pair,nogaki2022even,xu2024observation,strand2010transition,ran2019nearly,leggett1975theoretical,mackenzie2003superconductivity}.
Yet in microscopic lattice settings it is typically fragile: on-site attraction strongly favors singlet $s$-wave pairing and readily preempts odd-parity channels~\cite{read2000paired,wolf2022topological,wolf2022triplet,alicea2012new,beenakker2013search,stanescu2013majorana,elliott2015colloquium,Kitaev2001UnpairedMajorana}.
A key challenge is therefore to identify a minimal, controllable mechanism that both weakens the competing singlet channel and stabilizes an oriented triplet $p$-wave tendency, ideally in ultracold-atom optical lattices where state-dependent tunneling and the interaction range can be engineered with high precision~\cite{regal2003tuning,ticknor2004multiplet,zhang2004p,levinsen2007strongly,hu2019resonantly}.

Two ingredients naturally address this challenge. First, spin-dependent hopping anisotropy deforms the spin-resolved  Fermi surface  and suppresses conventional on-site $s$-wave coherence, opening phase space for unconventional pairing~\cite{feiguin2009exotic,gukelberger2014p}. Such Fermi surface geometry can also organize pairing correlations into a Cooper-pair Bose metal (CPBM), a gapless paired regime with a momentum-space "Bose surface" but without superfluid stiffness~\cite{phillips2003elusive,feiguin2009exotic,su2025variation,cao2024exotic}.
Second, a finite-range attraction---here taken as the nearest-neighbor (NN) attraction ($V<0$)---directly activates symmetry-allowed odd-parity channels and enhances triplet pairing tendencies once singlet pairing is weakened~\cite{nayak2021pairing,qu2022spin,cao2025dominant}.
However, how spin-dependent Fermi surface deformation and NN attraction cooperate to reshape both the strength and the orientation of the leading pairing correlations in a controlled lattice setting remains an open question.

In this work, we carry out constrained-path quantum Monte Carlo (CPQMC) simulations~\cite{zhang1997constrained,zhang1995constrained}
to study the attractive two-dimensional $t$–$U$–$V$ Hubbard model with the spin-dependent anisotropic hopping. 
At filling $n\simeq 0.85$, we map the ground-state phase diagram in the $(V,\alpha)$ plane, where $V$ denotes the NN attraction
and $\alpha$ parameterizes the spin-dependent hopping anisotropy. We identify three regimes:
an onsite $s$-wave superfluid ($s$-SF), a CPBM, and an oriented spin-triplet $p$-wave
pairing ($p$-TP) that appears for finite NN attraction strengths. 
Increasing the spin anisotropy suppresses the competing onsite $s$-wave channel and lowers the critical strength $|V_c(\alpha)|$ at which the $p$-wave vertex becomes dominant over the $s$-wave.
Momentum-resolved bubble-subtracted pairing correlators resolve an uncondensed Bose surface in the CPBM and a $\Gamma$-centered,
anisotropically elongated triplet vertex enhancement whose polar axis tracks the spin-resolved Fermi surface patches.
A channel-resolved Landau analysis provides a compact rationalization and yields a simple criterion for Landau scale $V^{L}_c(\alpha)$, 
whose anisotropy dependence tracks the numerical trend of $V_c(\alpha)$.

Motivated by cold-atom realizations of attractive Hubbard-type models with tunable interaction range~~\cite{Micnas1990superconductivity,randeria1989bound,Randeria1992PRL,Greiner2003Nature,Jochim2003Science}, 
we study the two-dimensional attractive $t$--$U$--$V$ Hubbard model on a square lattice with spin-dependent anisotropic hopping,
\begin{align}
H ={}& -\sum_{i,\sigma,\bm{\delta}} t_{\bm{\delta},\sigma} 
\left( \hat{c}^{\dagger}_{i,\sigma} \hat{c}_{i+\bm{\delta},\sigma} + \mathrm{h.c.} \right)
+ U \sum_{i} \hat{n}_{i\uparrow}\hat{n}_{i\downarrow}  \nonumber\\
&\quad + V \sum_{i,\bm{\delta},\sigma,\sigma'} 
\hat{n}_{i,\sigma}\hat{n}_{i+\bm{\delta},\sigma'}.
\label{eq:Hamiltonian}
\end{align}
Here $\hat{c}^\dagger_{i\sigma}$ ($\hat{c}_{i\sigma}$) creates (annihilates) a fermion with spin $\sigma=\uparrow,\downarrow$ on site $i$, and
$\hat{n}_{i\sigma}=\hat{c}^\dagger_{i\sigma}\hat{c}_{i\sigma}$ is the electron number operator. The hopping amplitude is $t_{\bm{\delta},\sigma
}$, where the displacement $\bm{\delta} \in \{ \hat{x}, \hat{y} \}$ connects site $i$ to its NN $i+\bm{\delta}$. Unless otherwise specified, we set $t \equiv 1$ and work at filling $n\simeq0.85$ with $U=-3$.   

\begin{figure}[!]
    \centering 
    \includegraphics[width=1\linewidth]{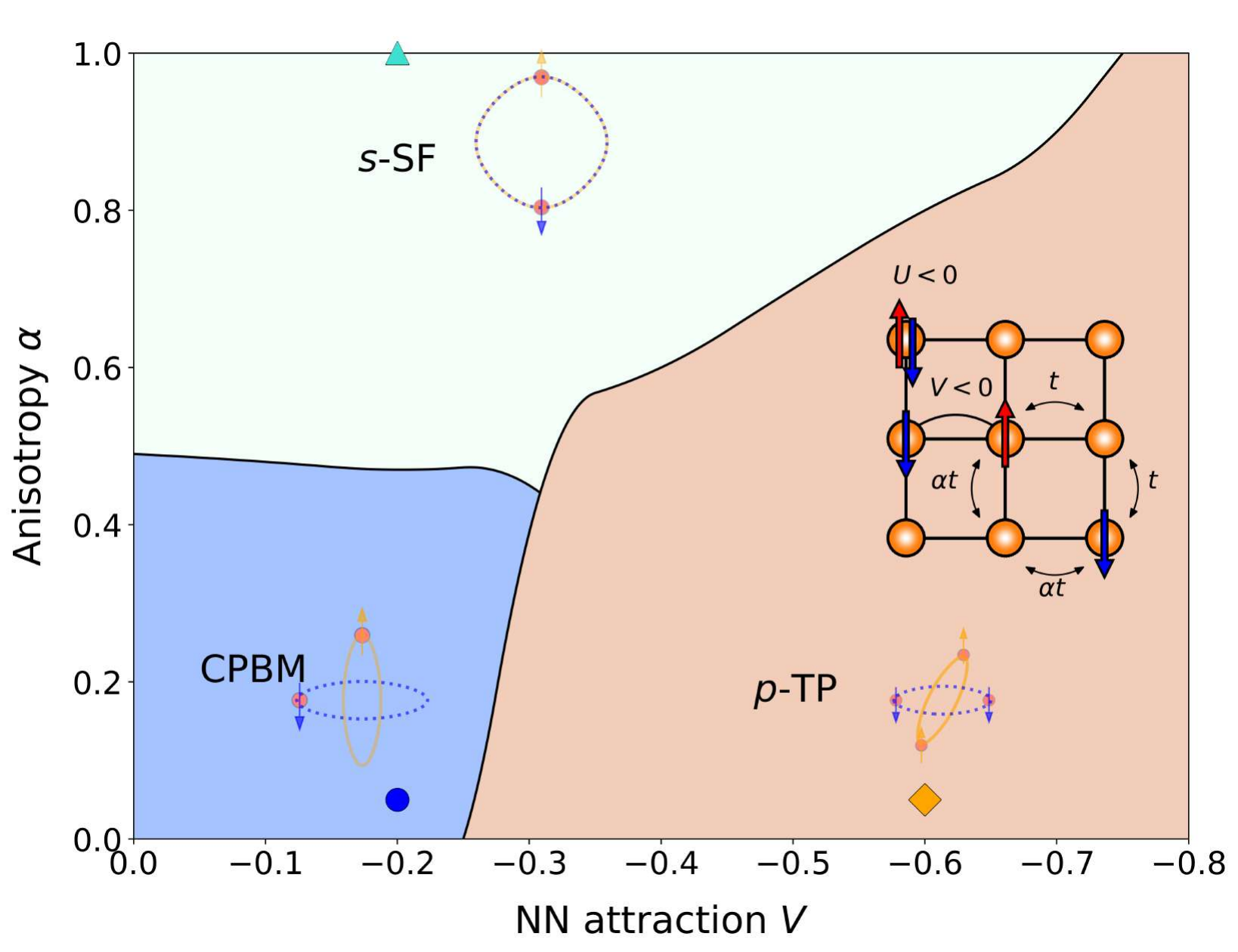}
    \caption{(Color online) Ground-state phase diagram of the spin-anisotropic attractive $t$--$U$--$V$ Hubbard model
[Eq.~\eqref{eq:Hamiltonian}.] at $U=-3$ (with $t=1$) and filling $n\simeq0.85$, estimated from quantum Monte Carlo simulations.
The horizontal axis is the nearest-neighbor attractive interaction $V$, and the vertical axis is the hopping
anisotropy parameter $\alpha$.
We identify an onsite $s$-wave superfluid ($s$-SF), a Cooper pair Bose metal (CPBM), and an oriented spin-triplet $p$-wave
pairing phase ($p$-TP). Decreasing $\alpha$ suppresses $s$-wave coherence and drives the system into the CPBM regime at small $|V|$, while a sufficiently
strong NN attraction, $|V|>|V_c(\alpha)|$, stabilizes the $p$-TP. Inset: spin-dependent hoppings
$t_{x\uparrow}=t_{y\downarrow}=t$ and $t_{y\uparrow}=t_{x\downarrow}=\alpha t$. Symbols denote representative data parameters in Fig.~\ref{fig3}.}
    \label{fig1}
\end{figure}

In addition to onsite attraction $U<0$, we include a NN attraction $V<0$ as the minimal short-range extension that captures the competition between conventional onsite singlet pairing and odd-parity pairing allowed by symmetry on the square lattice. As illustrated in the inset of Fig.~\ref{fig1}, we choose spin-dependent anisotropic NN hoppings
$t_{\hat{y},\downarrow}=t_{\hat{x},\uparrow}=t$ and $t_{\hat{x},\downarrow}=t_{\hat{y},\uparrow}=\alpha t$, and tune $\alpha\in[0,1]$ from the isotropic ($C_4$-symmetric) limit $\alpha=1$ to the quasi-one-dimensional limit $\alpha=0$; the hopping pattern is invariant under a $\pi/2$ lattice rotation followed by exchanging $\uparrow$ and $\downarrow$.
We work at zero spin polarization, $\langle \hat{n}_{i\uparrow}\rangle=\langle \hat{n}_{i\downarrow}\rangle=n/2$.
In this setting, $U$ favors onsite $s$-wave singlet correlations, whereas the finite-range attraction $V$ activates odd-parity $p$-wave channels.
For the spin-resolved Fermi-surface geometries relevant here, other even-parity NN form factors (extended-$s'$ and $d_{x^2-y^2}$) remain subleading (see Supplemental Material~\cite{tan2025-supp}); accordingly, we focus on the dominant $s$- and $p$-wave channels.

To diagnose these dominant pairing channels, we introduce the corresponding
equal-time pair operators: an onsite singlet, $\Delta_s(i)=c_{i\downarrow}c_{i\uparrow}$ and NN equal-spin triplets,
$\Delta_{p_{\eta},\sigma}(i)=\sum_{\bm{\delta}\in{\rm NN}}
f^{(\eta)}_{\bm{\delta}}\,c_{i+\bm{\delta},\sigma}c_{i\sigma}$,
with $\eta\in\{x,y\}$ and $\sigma\in\{\uparrow,\downarrow\}$.
The odd-parity form factors are subjected to the relation, 
$f^{(x/y)}_{+\hat{x}/\hat{y}}=+1$ and $f^{(x/y)}_{-\hat{x}/\hat{y}}=-1$, and vanish otherwise. 
We also define the equal-time single-particle Green's functions
$G^\sigma_{ij}\equiv \langle c^\dagger_{i\sigma}c_{j\sigma}\rangle$.

Following the bubble-subtracted (vertex) definition used in Ref.~\cite{cao2025dominant} (see also Refs.~\cite{white1989attractive,ying2018pairing,cao2024exotic}),
we remove the uncorrelated background built from one-body propagators and define the effective pair correlators as
\begin{align}
C^{\mathrm{eff}}_{s}(i,j)
&= \langle \Delta_s^{\dagger}(i)\Delta_s(j)\rangle
 -G^{\uparrow}_{i,j}G^{\downarrow}_{i,j},
\label{eq:ceff_s}
\\[2pt]
C^{\mathrm{eff},(\eta)}_{p,\sigma}(i,j)
&= \langle \Delta_{p_{\eta},\sigma}^{\dagger}(i)
          \Delta_{p_{\eta},\sigma}(j)\rangle
 -\Lambda^{(\eta)}_{p,\sigma}(i,j).
\label{eq:ceff_p}
\end{align}
with the bubble term
\begin{equation}
\Lambda^{(\eta)}_{p,\sigma}(i,j)
=\sum_{\bm{\delta},\bm{\delta}'\in\mathrm{NN}}
f^{(\eta)}_{\delta}\,f^{(\eta)}_{\delta'}\,
G^{\sigma}_{i,j}\,
G^{\sigma}_{i+\bm{\delta},\,j+\bm{\delta}'} .
\label{eq:Lambda_bubble}
\end{equation}

This subtraction isolates the interaction-driven vertex contribution in the Cooper channel,
so the effective vertex highlights the interaction-induced vertex enhancement in the Cooper channel that signals pairing.
Finally, we obtain the pairing momentum-resolved vertex distributions via
\begin{equation}
N^{\rm eff}_\ell(\mathbf{k})=\frac{1}{N}\sum_{i,j} e^{i\mathbf{k}\cdot(\mathbf{r}_i-\mathbf{r}_j)}\, C^{\rm eff}_\ell(i,j),
\qquad \ell\in\{s,(p_\eta,\sigma)\},
\end{equation}
where $N$ is the number of sites. For the triplet sector we report the dominant channel,
$N^{\rm eff}_{p\text{-pair}}(\mathbf{k})\equiv
\max_{\eta=x,y;\,\sigma=\uparrow,\downarrow}
N^{\rm eff}_{p_\eta,\sigma}(\mathbf{k})$,
and define its maximum value as $N^{\mathrm{eff}}_{p\text{-pair}}(\bm{k}_{\max}) \equiv \max_{\bm{k}} N^{\mathrm{eff}}_{p\text{-pair}}(\bm{k})$, where $\bm{k}_{\max}$ denotes the momentum at which $N^{\mathrm{eff}}_{p\text{-pair}}(\bm{k})$ is maximized. Details of the CPQMC implementation and parameter settings can be found in Ref.~\cite{cao2024exotic}.

Fig.~\ref{fig1} summarizes the ground-state phase diagram of the spin-anisotropic attractive $t$--$U$--$V$ Hubbard model on square lattices up to $20\times20$. 
The phase diagram is presented in the $(V,\alpha)$ plane, where phase boundaries are inferred from finite-size CPQMC data and serve as estimates for the thermodynamic limits. 
For weak anisotropy ($\alpha\simeq1$) and small $|V|$, the onsite attraction stabilizes $s$-SF. 
Upon decreasing $\alpha$, the spin-resolved Fermi surface deformation suppresses singlet coherence and eventually destroys the $s$-wave condensate, driving the system into the CPBM regime~\cite{su2025variation,cao2024exotic,phillips2003elusive,einstein2005quantentheorie,anderson1995observation,feiguin2009exotic}. At the vanishing NN attraction, we observe a crossover from $s$-SF to CPBM as $\alpha$ is reduced, while no signature of a dominant $p$-TP instability is found within the anisotropy range studied.

For $V<0$, the NN attraction provides an odd-parity pairing channel, and an oriented $p$-TP emerges once $|V|$ exceeds a critical value $|V_c(\alpha)|$ 
where the dominant $p$-wave vertex weight overtakes the $s$-wave vertex weight. Stronger anisotropies reduce $|V_c|$ by suppressing the competing onsite $s$-wave pairing correlations, thereby expanding the $p$-TP regime. In particular, at large spin anisotropies, the CPBM
found at small $|V|$ is replaced by the $p$-TP upon increasing $|V|$, highlighting how the spin anisotropy and nonlocal attraction 
cooperate to give rise to the $p$-TP: the spin anisotropy weakens $s$-wave coherence, while $V$ provides the odd-parity $p$-wave channel. 

We perform a minimal channel-resolved Landau analysis formulated in terms of normal- state Cooper bubbles~\cite{AltlandSimons2010,Tinkham1996,Thouless1960,NozieresSchmittRink1985,ChenLevin2005,nayak2018exotic,micnas1988superconductivity,hong2025unconventional}. Fourier transforming Eq.~(\ref{eq:Hamiltonian}) yields the spin-dependent dispersions
$\xi_{\mathbf{k}\uparrow}=-2t(\cos k_x+\alpha\cos k_y)-\mu$ and
$\xi_{\mathbf{k}\downarrow}=-2t(\alpha\cos k_x+\cos k_y)-\mu$, where $\mu$ is fixed by the filling. The momentum-dependent kernel function from the onsite and NN interaction reads 
\begin{equation}
V(\mathbf{k},\mathbf{k}')
 = U + 2V\bigl[\cos(k_x-k_x') + \cos(k_y-k_y')\bigr].
\label{eq:Vkk}
\end{equation}

Projecting $V(\mathbf k,\mathbf k')$ onto the symmetry-adapted basis $\{1,\sin k_x,\sin k_y\}$ for onsite $s$-wave and NN equal-spin 
$p$-wave pairing, we obtain the channel couplings $g_s = U$, $g_{p_x} = g_{p_y} = 2V$. 
The NN attraction also generates extended $s'$ and $d_{x^2-y^2}$ form factors, but their pairing susceptibilities are much smaller for Fermi surface geometries 
with spin anisotropy (see the Supplemental Material for details~\cite{tan2025-supp}).
Integrating out the fermions and keeping only quadratic terms in the pairing fields yields the Landau coefficients
\begin{align}
a_s(T,\alpha) &= \frac{1}{|U|} - \Pi_s(T,\alpha), \label{eq:a_s_main} \\
a_{p_\eta,\sigma}(T,\alpha) &= \frac{1}{2|V|} - \Pi_{p\eta,\sigma}(T,\alpha),
\quad \eta=x,y,\ \sigma=\uparrow,\downarrow .
\label{eq:a_p_main}
\end{align}
Here, $\Pi_s$ and $\Pi_{p\eta,\sigma}$ are the normal-state Cooper-channel susceptibilities (BCS bubbles)
computed from the free Green's functions
$G_{0,\sigma}(\mathbf{k},i\omega_n)=(i\omega_n-\xi_{\mathbf{k}\sigma})^{-1}$ with $\omega_n=(2n+1)\pi T$.
In practice, we evaluate these bubbles at low temperatures to reach the ground-state instabilities.
Performing the Matsubara sum yields the explicit forms
\begin{align}
&\Pi_s(T,\alpha) = \frac{1}{N}\sum_{\mathbf{k}}
\frac{\tanh(\xi_{\mathbf{k}\uparrow}/2T)+\tanh(\xi_{\mathbf{k}\downarrow}/2T)}
     {2(\xi_{\mathbf{k}\uparrow}+\xi_{\mathbf{k}\downarrow})},\\
&\Pi_{p\eta,\sigma}(T,\alpha) = \frac{1}{N}\sum_{\mathbf{k}}
\sin^2 k_\eta\,\frac{\tanh(\xi_{\mathbf{k}\sigma}/2T)}{2\xi_{\mathbf{k}\sigma}},
\end{align}
whose derivation and bubble diagrams are provided in the Supplemental Material~\cite{tan2025-supp}.

\begin{figure}[t!]
    \centering 
    \includegraphics[width=1\linewidth]{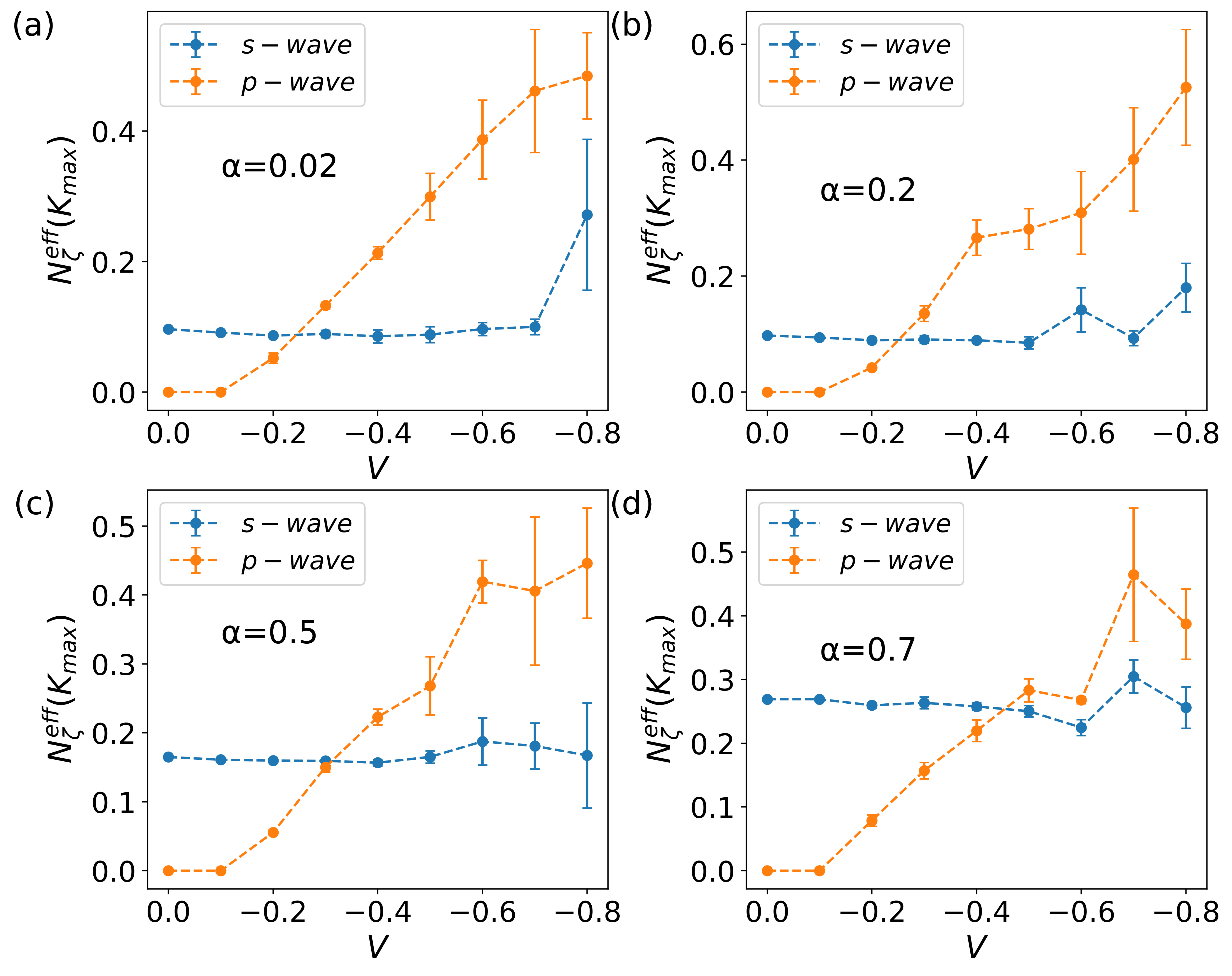}
    \caption{(Color online) Maximum vertex weight
$N^{\rm eff}_\ell(\mathbf{k}_{\max})$
for the onsite $s$-wave (blue) and NN equal-spin $p$-wave (orange) channels versus NN attraction $V$.
Panels (a)--(d) correspond to hopping anisotropy $\alpha=0.02$, $0.2$, $0.5$, and $0.7$, respectively.
The error bars denote statistical uncertainties.}

    \label{fig2}
\end{figure}

Because $\Pi_s$ involves both spin branches, decreasing $\alpha$ reduces $\Pi_s$ via Fermi surface mismatch and thus suppresses onsite $s$-wave coherence.
In contrast, $\Pi_{p_\eta,\sigma}$ is an equal-spin bubble on a single branch weighted by the odd-parity form factor $\sin^2k_\eta$. 
Therefore, for each spin, anisotropy lifts the near-degeneracy between the $p_x$ and $p_y$ components and the dominant triplet sector 
is selected via $\max_{\eta,\sigma}\Pi_{p_\eta,\sigma}(T,\alpha)$. 
At the Gaussian level, the Thouless condition $a_{p_\eta,\sigma}(T,\alpha)=0$ defines a Landau scale for the onset of the leading triplet instability,
\begin{equation}
|V_c^{\rm L}(\alpha)|=\Big[2\max_{\eta,\sigma}\Pi_{p_\eta,\sigma}(T,\alpha)\Big]^{-1}.
\label{eq:VcL}
\end{equation}

To compactly compare the pairing tendencies across channels, we summarize the momentum-resolved vertex
distributions $N^{\rm eff}_\ell(\mathbf{k})$ by their maxima
$N^{\rm eff}_\ell(\mathbf{k}_{\max})$.
Fig.~\ref{fig2} shows $N^{\rm eff}_s(\mathbf{k}_{\max})$ and the dominant NN equal-spin $p$-wave signal as functions of $V$ for several anisotropies $\alpha$.
 For strong spin-dependent hopping anisotropy ($\alpha=0.02$), $N^{\rm eff}_s(\mathbf{k}_{\max})$ depends only weakly on $V$,
consistent with the onsite attraction $U<0$ controlling the singlet channel,
whereas the $p$-wave vertex weight drops rapidly toward the noise floor as $V\!\to\!0^{-}$.
The crossing of the two curves provides an empirical estimate of the critical attraction $V_c(\alpha)$ beyond which
the $p$-wave vertex becomes dominant (e.g., $V_c(0.02)\simeq -0.23$).
Upon increasing $\alpha$ [Figs.~\ref{fig2}(b)--~\ref{fig2}(d)], the crossing shifts systematically to more negative $V$,
indicating that reduced anisotropy requires stronger NN attraction to stabilize dominant $p$-wave correlations.
This trend mirrors the decrease of the Landau $p$-wave scale
$|V_c^{\mathrm L}(\alpha)|=[2\max_{\eta,\sigma}\Pi_{p_\eta,\sigma}(T,\alpha)]^{-1}$ in Eq.~\eqref{eq:VcL}.

\begin{figure}[t!]
    \centering 
    \includegraphics[width=1\linewidth]{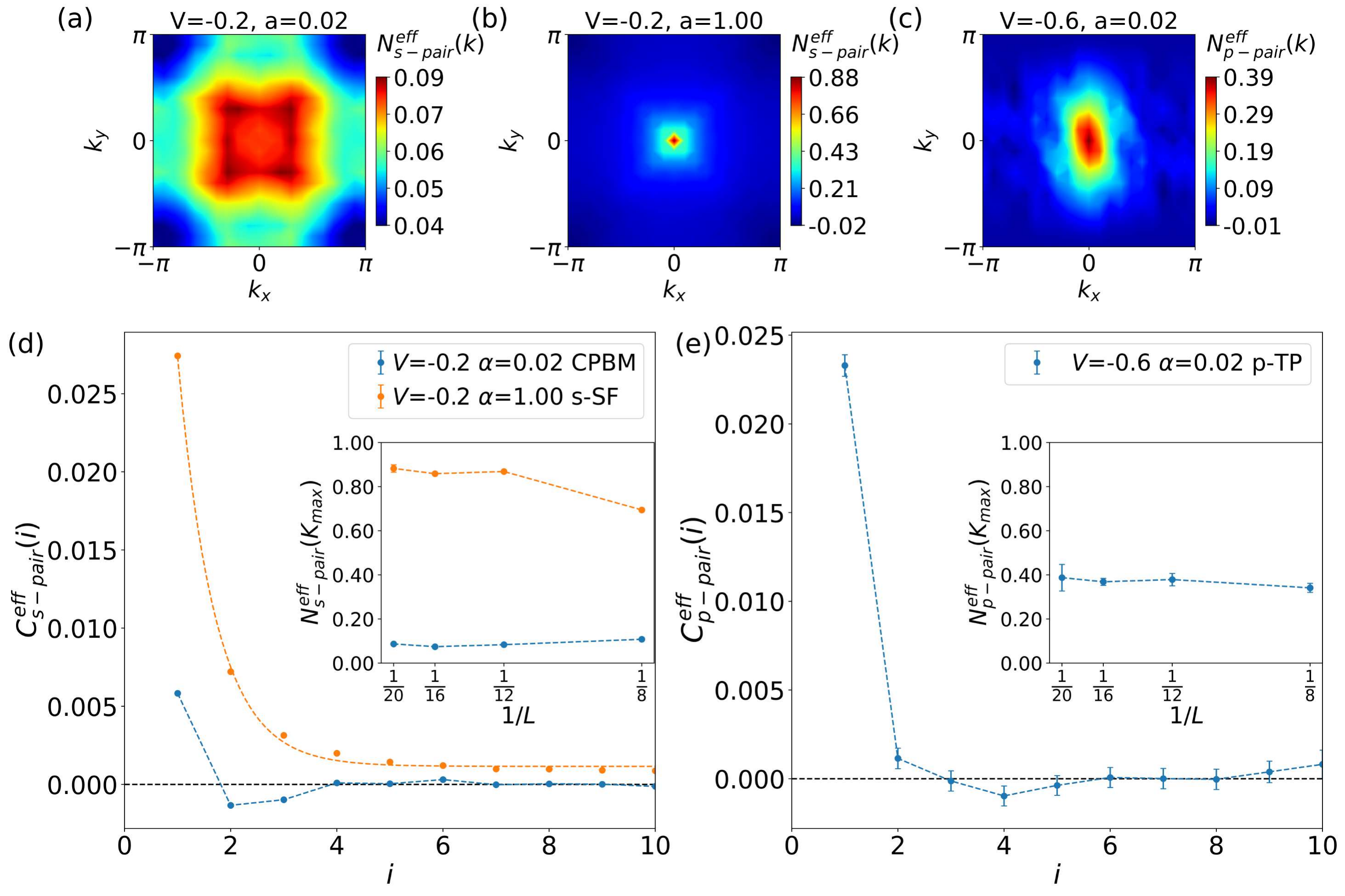}
    \caption{(Color online) Pairing vertex distributions and correlators in the $t$--$U$--$V$ model.
Panels (a)--(c) correspond to the CPBM (circle), $s$-SF (triangle), and p-TP (diamond) points in Fig.~\ref{fig1}:
(a) $N^{\rm eff}_{s\text{-pair}}(\mathbf{k})$ at $V=-0.2$, $\alpha=0.02$;
(b) $N^{\rm eff}_{s\text{-pair}}(\mathbf{k})$ at $V=-0.2$, $\alpha=1.00$;
(c) $N^{\rm eff}_{p\text{-pair}}(\mathbf{k})$ at $V=-0.6$, $\alpha=0.02$.
(d) $C^{\rm eff}_{s\text{-pair}}(i)$ for (a) and (b); inset: scaling of $N^{\rm eff}_{s\text{-pair}}(k_{\max})$.
(e) $C^{\rm eff}_{p\text{-pair}}(i)$ for (c); inset: scaling of $N^{\rm eff}_{p\text{-pair}}(k_{\max})$.}
    \label{fig3}
\end{figure}

Having established from Figs.~\ref{fig1} and \ref{fig2} that the leading pairing channel switches between the $s$- and $p$-wave sectors as $(V,\alpha)$ are tuned, we now examine momentum- and real-space signatures of the effective pairing correlations. In the CPBM [Fig.~\ref{fig3}(a)], $N^{\mathrm{eff}}_{s\text{-pair}}(\mathbf{k})$ forms a broad, continuous nonzero  momentum Bose surface, consistent with preformed yet uncondensed Cooper pairs and the absence of global phase coherence~\cite{feiguin2009exotic}.
By contrast, in the $s$-SF [Fig.~\ref{fig3}(b)] the distribution develops a sharp peak at $\mathbf{k}=0$ , signaling Bose condensation of Cooper pairs and long-range phase coherence.
The real-space correlator $C^{\mathrm{eff}}_{s\text{-pair}}(i)$ [Fig.~\ref{fig3}(d)] decays rapidly and oscillates about zero in the CPBM, but remains positive over much longer distances in the $s$-SF.
Consistently, the inset shows that $N^{\mathrm{eff}}_{s\text{-pair}}(k_{\max})$ extrapolates to a finite value in the $s$-SF, whereas in the CPBM it saturates with system size, consistent with the absence of off-diagonal long range order.

Now we focus on the $p$-TP regime and show the effective $p$-wave momentum distribution $N^{\mathrm{eff}}_{p\text{-pair}}(\mathbf{k})$ at strong anisotropy $\alpha=0.02$ and $V=-0.6$, as shown in Fig.~\ref{fig3}(c).
A pronounced lobe centered at $\Gamma$ elongates along a single momentum axis, indicating a polar selection between the nearly degenerate $p_x/p_y$ components.
This oriented selection reflects the anisotropy-induced reshaping of the spin-resolved Fermi surface patches and the resulting lifting of the $p_x$--$p_y$ near-degeneracy, consistent with our Landau analysis.
In real-space [Fig.~\ref{fig3}(e)], $C^{\mathrm{eff}}_{p\text{-pair}}(i)$ remains small and decays rapidly over accessible sizes, so establishing asymptotic off-diagonal long range order from $C^{\mathrm{eff}}_{p\text{-pair}}(r)$ alone is challenging.
Nevertheless, the inset shows that the maximal vertex weight $N^{\mathrm{eff}}_{p\text{-pair}}(k_{\max})$ remains appreciable and exhibits a weak but systematic size dependence, suggesting enhanced interaction-driven triplet pairing correlations in this regime.

\begin{figure*}[]
\centering
\includegraphics[width=\linewidth]{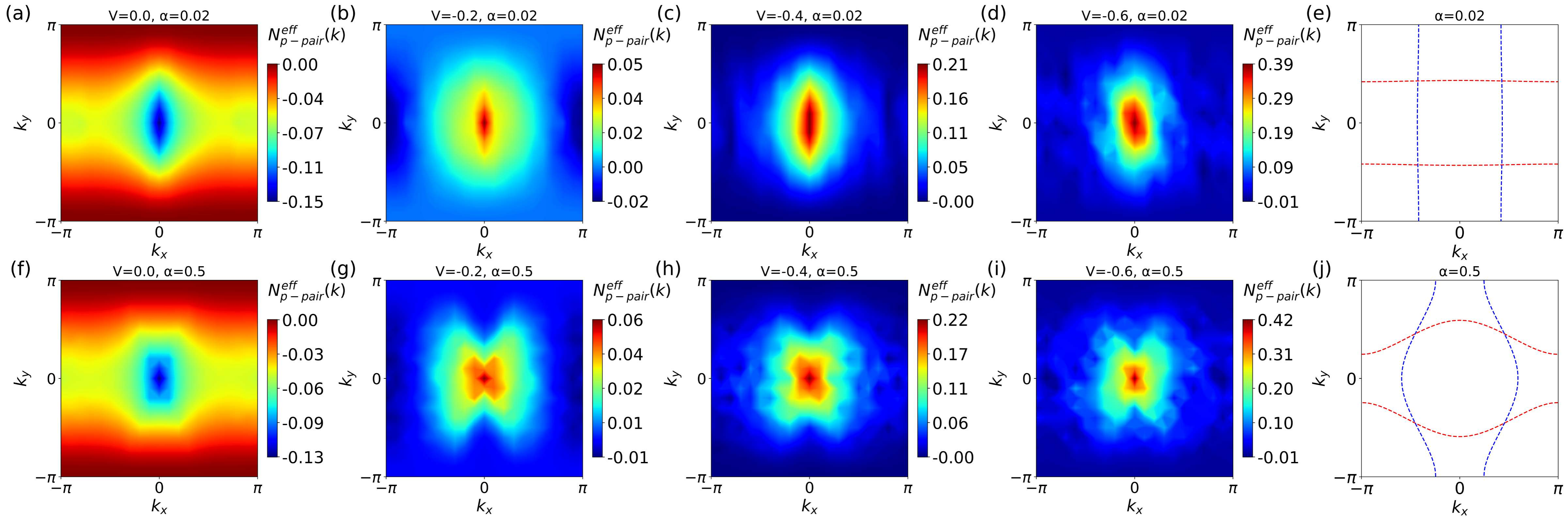}
\caption{(Color online)
Momentum-space evolution of $N^{\mathrm{eff}}_{p\text{-pair}}(\mathbf{k})$.
Top row: $\alpha = 0.02$.  
Panels (a)–(d) correspond to $V = 0, -0.2, -0.4, -0.6$, and panel (e) shows
the associated spin-resolved Fermi surface for $\uparrow/\downarrow$ 
(blue/red dashed curves).  
Bottom row: $\alpha = 0.50$.  
Panels (f)–(i) show the same sequence of $V$, and panel (j) displays the 
corresponding Fermi surface.  
For $V=0$, the $p$-wave signal is absent; increasing $|V|$ 
redistributes spectral weight toward $\Gamma$ and produces oriented 
lobes signaling the emergence of odd-parity triplet correlations.  
For stronger anisotropy (smaller $\alpha$), the distribution elongates 
along a single polar axis ($p_x$ or $p_y$), indicating that anisotropy 
selects the orientation of the $p$-wave channel via Fermi surface deformation.  
At fixed $V$, the total magnitude of the $p$-wave signal remains comparable 
between the two anisotropies, with only minor numerical variations.
}
\label{fig4}
\end{figure*}

To expose how the polar axis of the $p$-TP phase is selected microscopically, we next examine in Fig.~\ref{fig4} the full momentum-space evolution of the $p$-wave vertex distribution $N^{\rm eff}_{p\text{-pair}}(\mathbf{k})$ as a function of the NN attraction $V$ and the hopping anisotropy $\alpha$.
Panels~\ref{fig4}(a)–\ref{fig4}(d) show a sequence of $V$ at strong anisotropy $\alpha=0.02$, while panels~\ref{fig4}(f)–\ref{fig4}(i) display the same $V$ values at moderate anisotropy $\alpha=0.50$.
The rightmost panels~\ref{fig4}(e) and \ref{fig4}(j) plot the corresponding spin-resolved Fermi surface, highlighting how spin-dependent hopping reshapes the spin-split Fermi surface patches.
Because $N^{\rm eff}_{p\text{-pair}}(\mathbf{k})$ is bubble-subtracted, it fluctuates around zero ($p$-wave pairing is entirely absent) when $V=0$, providing a clean baseline for isolating interaction-induced vertex enhancement.

At fixed $\alpha$, increasing $|V|$ immediately generates a pronounced odd-parity vertex contribution concentrated near $\Gamma$, and the overall scale of $N^{\rm eff}_{p\text{-pair}}(\mathbf{k})$ grows with $|V|$.
This directly visualizes the mechanism captured by the projected coupling $g_{p_x}=g_{p_y}=2V$ and the channel-resolved Landau analysis: nonlocal attraction activates the equal-spin triplet Cooper channel, while the momentum-space vertex distribution reveals where the enhancement is concentrated.

In contrast, the anisotropy parameter $\alpha$ primarily regulates the structure and orientation of the $p$-wave correlations.
At the strong anisotropy $\alpha=0.02$, the weight near $\Gamma$ is sharply elongated along a single momentum axis [Figs.~\ref{fig4}(a)–\ref{fig4}(d)], consistent with strongly distorted spin-resolved Fermi surface patches [Fig.~\ref{fig4}(e)] that lift the near-degeneracy between the $p_x$ and $p_y$ components.
For $\alpha=0.50$, the spin-resolved Fermi surface becomes more nearly $C_4$ symmetric [Fig.~\ref{fig4}(j)], and the $p$-wave pairing distribution correspondingly evolves toward a more fourfold, quasi-isotropic lobe structure [Figs.~\ref{fig4}(f)–\ref{fig4}(i)].
Thus, within the regime studied, anisotropy does not primarily amplify the total triplet weight; rather, it selects the polar orientation and reshapes the momentum-space morphology of the $p$-wave pairing correlations through Fermi surface deformation.

Taken together, the results establish a coherent microscopic picture for the emergence of an oriented triplet state in the spin-anisotropic attractive $t$--$U$--$V$ Hubbard model. 
A finite NN attraction is essential to activate an odd-parity Cooper channel, while the spin-dependent hopping anisotropy $\alpha$ suppresses the competing onsite singlet coherence and, by deforming the spin-resolved Fermi surface, selects a polar axis (either $p_x$ or $p_y$) for the triplet correlations. 
This cooperative mechanism explains both the topology of the $(V,\alpha)$ phase diagram and the strongly anisotropic momentum-space vertex patterns in the $p$-TP regime.

In summary, we mapped the ground-state phase diagram of the attractive $t-U-V$ Hubbard model at $n\simeq0.85$, identified CPBM, $s$-SF
and $p$-TP phases using quantum Monte Carlo simulations. Momentum-resolved vertex distributions and real-space correlators provide distinct signatures: the CPBM exhibits a broad Bose-surface feature without a condensate peak and no off-diagonal long-range order, the $s$-SF develops a sharp $\Gamma$-point condensate peak, and the $p$-TP shows an anisotropically elongated $p$-wave weight locked to the spin-split Fermi surface patches. 
Finally, a channel-resolved Landau analysis yields a compact $p$-wave Thouless scale $|V_c^{\mathrm L}(\alpha)|$ [Eq.~\eqref{eq:VcL}], whose anisotropy dependence tracks the crossover boundary into the triplet dominant regime.

More broadly, the mechanism uncovered here---finite-range attraction opening an odd-parity channel and momentum-space Fermi surface geometry selecting its orientation---should be relevant in other spin-split metals. 
In particular, altermagnetic band structures exhibit symmetry-enforced momentum-dependent spin polarization without net magnetization, producing spin-resolved Fermi surface patches analogous to those controlling the present instability~\cite{fedchenko2024trsb_ruo2,krempasky2024kramers,osumi2024mnte,reimers2024crsb_natcomm,zeng2024crsb_advsci,ding2024gwave_crsb_prl,noda2016mno2_pccp,smejkal2020scadv_hall,hayami2019jpsj_spinsplitting,ahn2019ruo2_pomeranchuk,yuan2020lowz_prb,yuan2021prm_prediction,smejkal2022prx_landscape,smejkal2022prx_beyond,mazin2022prx_editorial,bai2024afm_review}. 
Therefore, it would be interesting to extend the present microscopic framework to other altermagnetic Hamiltonians with effective attractive interactions, 
and further identify the finite-momentum paired state.

Z.C. acknowledges support from China Postdoctoral Science Foundation (2025M783397). H.K.T acknowledges support from  the Shenzhen Fundamental Research Program (No.~JCYJ20250604145655074) and Shenzhen Key Laboratory of Advanced Functional Carbon Materials Research and Comprehensive Application~(No.~ZDSYS20220527171407017). T.Y. acknowledges support from Natural Science Foundation of Heilongjiang Province~(No.~YQ2023A004). M.Z. is supported by the National Natural Science Foundation of China under Grant No. 12504172. This work is supported by the Research Foundation of Yanshan University under Grant No.~8190448.

The data are available from the authors upon reasonable request.
\bibliography{ref}
\end{document}


\title{Supplemental Material for ``Oriented Triplet $p$-Wave Pairing from Fermi surface Anisotropy and Nonlocal Attraction"}

\author{Shuning Tan}
\thanks{These authors contributed equally.}
\affiliation{Key Laboratory for Microstructural Material Physics of Hebei Province, School of Science, Yanshan University, Qinhuangdao 066004, China} 

\author{Ji Liu}
\thanks{These authors contributed equally.}
\affiliation{School of Science, Harbin Institute of Technology, Shenzhen, 518055, China}
\affiliation{Shenzhen Key Laboratory of Advanced Functional Carbon Materials Research and Comprehensive Application, Shenzhen 518055, China.}

\author{Minghuan Zeng}
\affiliation{College of Physics, Chongqing University, Chongqing 401331, China}

\author{Tao Ying}
\affiliation{School of Physics, Harbin Institute of Technology, Harbin 150001, China}

\author{Zhangkai Cao}
\email{caozhangkai@mail.bnu.edu.cn}
\affiliation{Eastern Institute for Advanced Study, Eastern Institute of Technology, Ningbo, Zhejiang 315200, China}
\affiliation{School of Physical Sciences, University of Science and Technology of China, Hefei, 230026, China}

\author{Ho-Kin Tang}
\email{denghaojian@hit.edu.cn}
\affiliation{School of Science, Harbin Institute of Technology, Shenzhen, 518055, China}
\affiliation{Shenzhen Key Laboratory of Advanced Functional Carbon Materials Research and Comprehensive Application, Shenzhen 518055, China.}

\date{\today} 
\maketitle 

In this Supplemental Material, we provide additional analytical details and numerical results supporting the main text.
In Sec.~S1, we derive the quadratic Landau coefficients used in Eqs.(6)--(11) of the main text by projecting the
microscopic interactions onto symmetry-adapted pairing channels and evaluating the corresponding normal-state
pairing susceptibilities.
We also present CPQMC momentum-resolved vertex distributions for the extended-$s'$ and $d_{x^2-y^2}$ channels
to confirm that they remain subleading in the parameter regime studied.
In Sec.~S2, we provide additional CPQMC results for the momentum-resolved effective $s$- and $p$-wave pair
distributions $N^{\mathrm{eff}}_{\ell}(\mathbf{k})$, illustrating the anisotropy-driven evolution across the $s$-SF, CPBM, and $p$-TP regimes.

\section*{S1. Mean-field origin of the Landau coefficients}
For completeness, we outline the minimal steps leading from the microscopic interactions to the quadratic Landau
coefficients quoted in the main text.

\subsection*{A. Projection of the microscopic interactions}
For later convenience, we rewrite the spin-anisotropic attractive $t$--$U$--$V$ Hubbard model of Eq.~(1) of the main text
in momentum space and separate it into a quadratic part and an interaction part,
\begin{equation}
\hat H=\hat H_{0}+\hat H_{\mathrm{int}} ,
\end{equation}
with
\begin{align}
\hat H_{0} &= \sum_{\mathbf{k},\sigma}\xi_{\mathbf{k}\sigma}\,
\hat c^{\dagger}_{\mathbf{k}\sigma}\hat c_{\mathbf{k}\sigma}, \\
\hat H_{\mathrm{int}} &= \frac{1}{N}\sum_{\mathbf{k},\mathbf{k}'}\sum_{\sigma,\sigma'}
V(\mathbf{k},\mathbf{k}')\,
\hat c^{\dagger}_{\mathbf{k}\sigma}\hat c^{\dagger}_{-\mathbf{k}\sigma'}
\hat c_{-\mathbf{k}'\sigma'}\hat c_{\mathbf{k}'\sigma}.
\end{align}
Here $\xi_{\mathbf{k}\sigma}$ is the spin-dependent anisotropic dispersion given in the main text, and
$V(\mathbf{k},\mathbf{k}')$ is the interaction kernel defined in Eq.~(6).
Throughout we focus on uniform pairing instabilities and hence restrict to the zero center-of-mass ($\mathbf{q}=0$) Cooper channel.

Introducing the symmetry-adapted basis functions
\begin{equation}
\label{eq:formfactor}
\begin{aligned}
\phi_s(\mathbf{k}) &= 1, \qquad
\phi_{s'}(\mathbf{k}) = \frac{1}{2}\left(\cos k_x + \cos k_y\right), \qquad
\phi_{d_{x^2-y^2}}(\mathbf{k}) = \frac{1}{2}\left(\cos k_x - \cos k_y\right),\\
\phi_{p_x}(\mathbf{k}) &= \sin k_x, \qquad
\phi_{p_y}(\mathbf{k}) = \sin k_y .
\end{aligned}
\end{equation}
projecting $V(\mathbf{k},\mathbf{k}')$ onto these channels yields the bare channel couplings
\begin{equation}
g_s=U,\qquad g_{s'}=4V,\qquad g_{d_{x^2-y^2}}=4V,\qquad g_{p_x}=g_{p_y}=2V.
\end{equation}
For $U<0$ and $V<0$, the channel couplings $g_\ell$ are negative. 
It is therefore convenient to introduce the attraction strengths
$|g_\ell|\equiv -g_\ell>0$ and write the interaction projected onto the
symmetry-adapted Cooper channels in separable form,
\begin{align}
\hat H_{\rm int}
= {}&
-\sum_{\ell\in\{s,s',d_{x^2-y^2}\}}\frac{|g_\ell|}{N}
\sum_{\mathbf{k},\mathbf{k}'}
\phi_\ell(\mathbf{k})\phi_\ell(\mathbf{k}')
\hat c^\dagger_{\mathbf{k}\uparrow}\hat c^\dagger_{-\mathbf{k}\downarrow}
\hat c_{-\mathbf{k}'\downarrow}\hat c_{\mathbf{k}'\uparrow}
\nonumber\\
&\quad
-\sum_{\eta=x,y}\sum_{\sigma=\uparrow,\downarrow}\frac{|g_{p_\eta}|}{N}
\sum_{\mathbf{k},\mathbf{k}'}
\phi_{p_\eta}(\mathbf{k})\phi_{p_\eta}(\mathbf{k}')
\hat c^\dagger_{\mathbf{k}\sigma}\hat c^\dagger_{-\mathbf{k}\sigma}
\hat c_{-\mathbf{k}'\sigma}\hat c_{\mathbf{k}'\sigma}.
\label{eq:S7_new}
\end{align}

Although the NN attraction generates identical bare couplings $g_{s'}=g_{d_{x^2-y^2}}=4V$ in the extended-$s'$ and
$d_{x^2-y^2}$ channels, their impact on pairing is controlled by the corresponding susceptibilities.
As a consistency check, we compute the momentum-resolved effective pairing weights in these channels using CPQMC
(see Figs.~S1 and S2), which confirms that the extended-$s'$ and $d_{x^2-y^2}$ channels remain subleading in the
parameter regime of interest.

\begin{figure*}[htb!]
    \centering
    \includegraphics[width=0.7\linewidth]{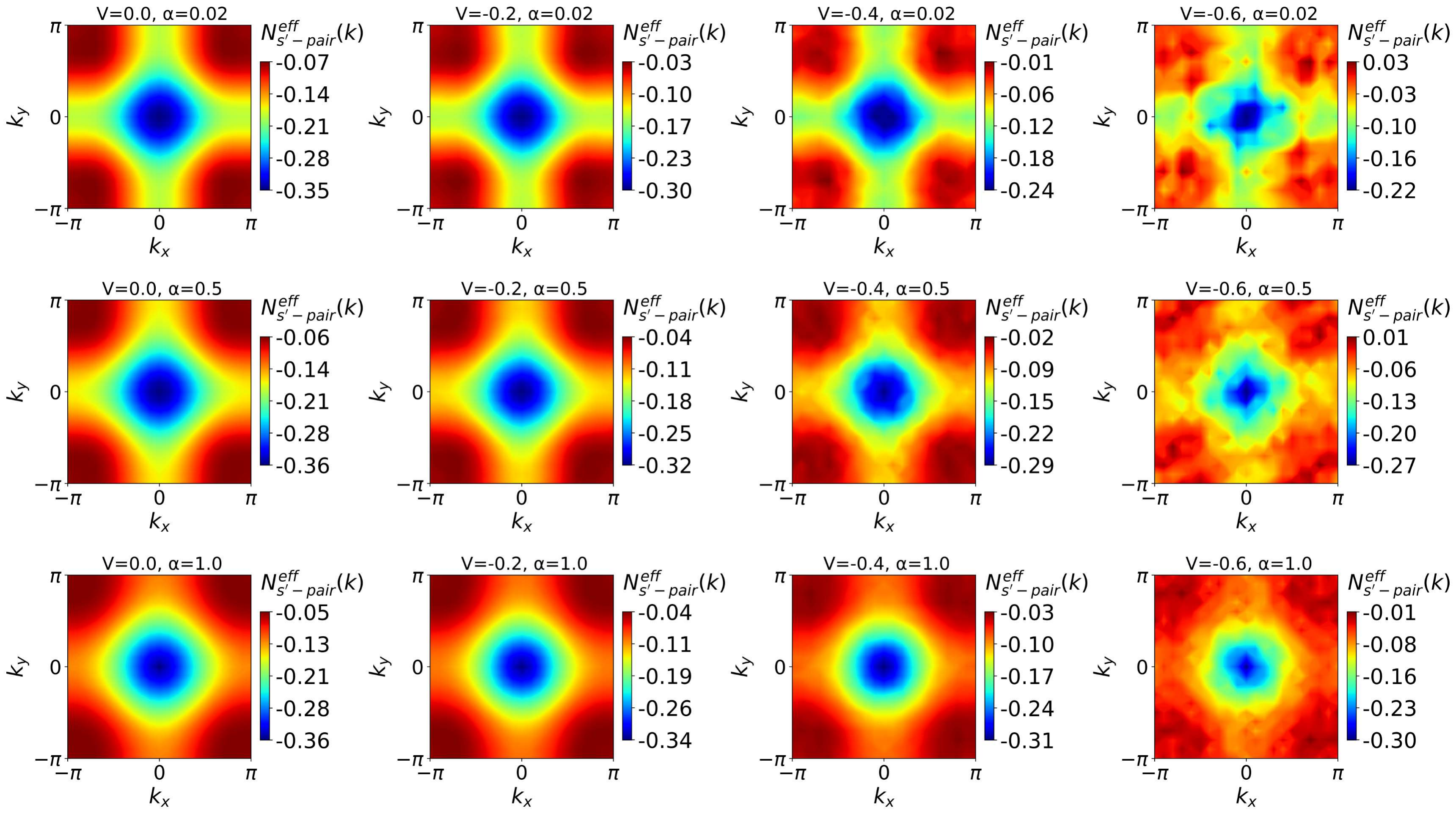}
    \caption{(Color online) CPQMC results for the effective extended-$s$-wave
pair momentum distribution $N^{\rm eff}_{s'\text{-pair}}(\mathbf{k})$ on a
$20\times20$ lattice at filling $n\simeq0.85$, shown for several values of
the nearest-neighbor attraction $V$ and hopping anisotropy $\alpha$.
Although the extended-$s$ channel is symmetry-allowed by the NN attraction,
its overall magnitude remains small and no sharp condensation features
develop, indicating that this channel is subleading throughout the
parameter regime studied.
}
\label{figS1}
\end{figure*}

\begin{figure*}[htb!]
    \centering
    \includegraphics[width=0.7\linewidth]{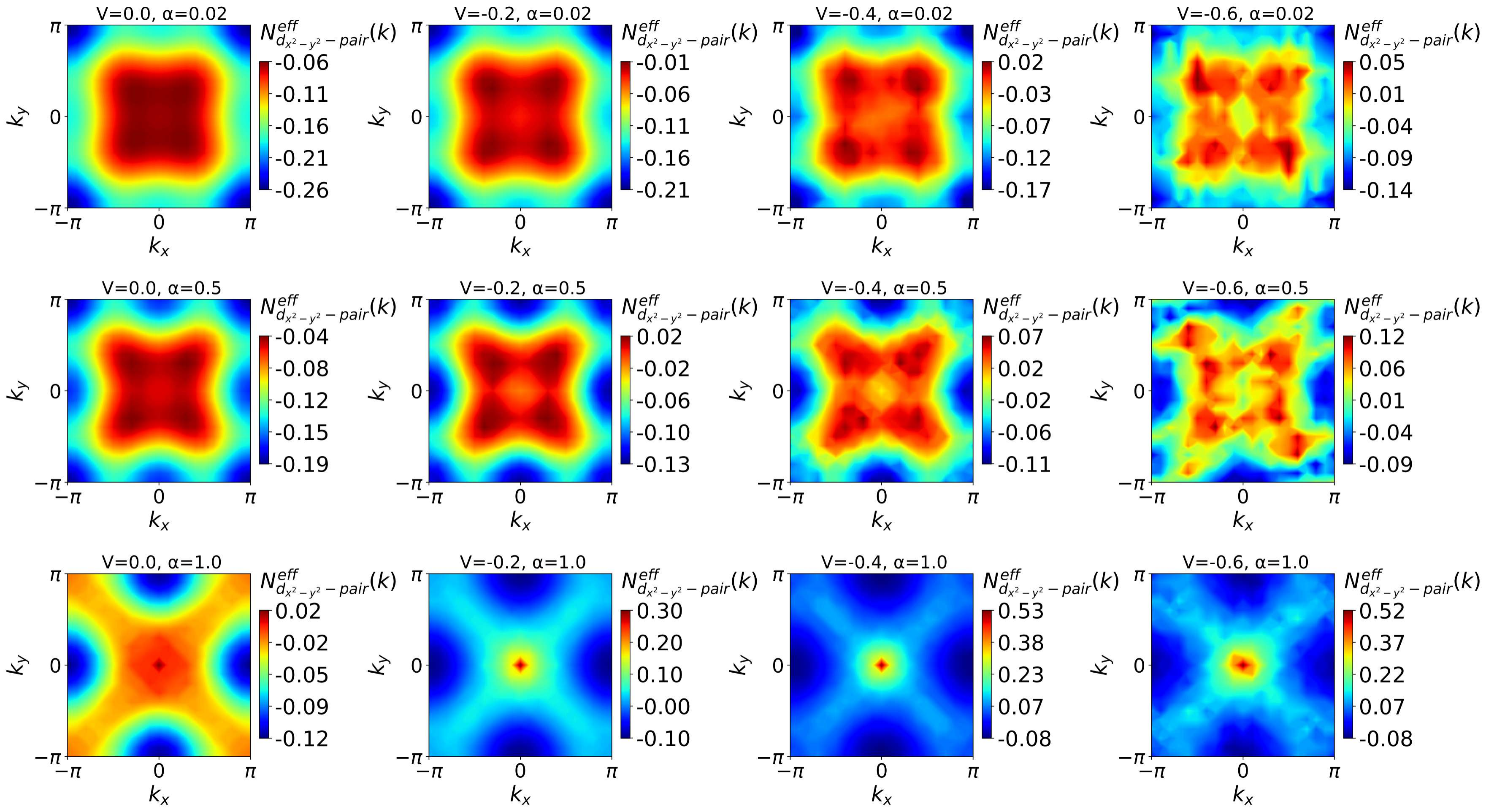}
    \caption{(Color online) CPQMC results for the effective
$d_{x^2-y^2}$-wave pair momentum distribution
$N^{\rm eff}_{d_{x^2-y^2}\text{-pair}}(\mathbf{k})$ on a $20\times20$ lattice
at filling $n\simeq0.85$, for the same set of parameters as in Fig.~\ref{figS1}.
Despite the presence of a finite projected coupling $g_{d_{x^2-y^2}}=4V$,
the $d$-wave channel remains weak and does not exhibit a dominant pairing
instability, consistent with the channel-resolved Landau analysis 
}
    \label{figS2}
\end{figure*}

\subsection*{B. Grassmann path-integral representation}

We start from the partition function $Z=\mathrm{Tr}\,e^{-\beta H}$ and use fermionic coherent states to write it as a Grassmann path integral~\cite{sigrist1991phenomenological,mineev1999introduction,samokhin2024ginzburg,blount1985symmetry,dalal2023field,honerkamp2004bcs},
\begin{equation}
Z=\int \mathcal{D}[\bar{\psi},\psi]\; e^{-S[\bar{\psi},\psi]},
\end{equation}
with the Euclidean action
\begin{align}
S[\bar\psi,\psi]
= \int_0^\beta d\tau \Bigg[
&\sum_{\mathbf{k},\sigma}\bar\psi_{\mathbf{k}\sigma}(\tau)
\bigl(\partial_\tau+\xi_{\mathbf{k}\sigma}\bigr)\psi_{\mathbf{k}\sigma}(\tau)
\nonumber\\
&-\sum_{\ell\in\{s,s',d_{x^2-y^2}\}}\frac{|g_\ell|}{N}
\sum_{\mathbf{k},\mathbf{k}'}
\phi_\ell(\mathbf{k})\phi_\ell(\mathbf{k}')
\bar\psi_{\mathbf{k}\uparrow}(\tau)\bar\psi_{-\mathbf{k}\downarrow}(\tau)
\psi_{-\mathbf{k}'\downarrow}(\tau)\psi_{\mathbf{k}'\uparrow}(\tau)
\nonumber\\
&-\sum_{\eta=x,y}\sum_{\sigma=\uparrow,\downarrow}\frac{|g_{p_\eta}|}{N}
\sum_{\mathbf{k},\mathbf{k}'}
\phi_{p_\eta}(\mathbf{k})\phi_{p_\eta}(\mathbf{k}')
\bar\psi_{\mathbf{k}\sigma}(\tau)\bar\psi_{-\mathbf{k}\sigma}(\tau)
\psi_{-\mathbf{k}'\sigma}(\tau)\psi_{\mathbf{k}'\sigma}(\tau)
\Bigg].
\label{eq:S11_new}
\end{align}
Here $\bar\psi_{\mathbf{k}\sigma}(\tau)$ and $\psi_{\mathbf{k}\sigma}(\tau)$ are Grassmann fields,
$\beta=1/T$.
The form factors $\phi_\ell(\mathbf{k})$ are defined in Eq.~\ref{eq:formfactor}.
We restrict to uniform ($\mathbf{q}=0$) pairing instabilities and thus keep only the zero-momentum Cooper channel.

\subsection*{C. Hubbard--Stratonovich decoupling in pairing channels}

The quartic interactions in the projected ($\mathbf{q}=0$) Cooper channels can be decoupled by introducing
complex Hubbard--Stratonovich (HS) pairing fields~\cite{sigrist1991phenomenological,mineev1999introduction,samokhin2024ginzburg,blount1985symmetry,dalal2023field,honerkamp2004bcs}. This HS transformation is an \emph{exact} identity; a
mean-field approximation is invoked only if one subsequently evaluates the HS functional integral at a saddle point.

In this work we keep only the on-site singlet $s$ channel and the nearest-neighbor (NN) equal-spin triplet $p$ channels.
The corresponding separable form factors are
$\phi_s(\mathbf{k})=1$ and $\phi_{p\eta}(\mathbf{k})=\sin k_\eta$ with $\eta=x,y$ (so that
$\phi_{p\eta}(-\mathbf{k})=-\phi_{p\eta}(\mathbf{k})$, ensuring the required antisymmetry for equal-spin pairing).
The attraction strengths are $|g_s|=|U|$ and $|g_p|=2|V|$.

\paragraph*{Singlet $s$ channel.}
The on-site Cooper interaction can be written as
\begin{align}
\exp\Bigg[
\frac{|U|}{N}\int_0^\beta d\tau
\sum_{\mathbf{k},\mathbf{k}'}
\bar\psi_{\mathbf{k}\uparrow}(\tau)\bar\psi_{-\mathbf{k}\downarrow}(\tau)\,
\psi_{-\mathbf{k}'\downarrow}(\tau)\psi_{\mathbf{k}'\uparrow}(\tau)
\Bigg]
\propto
\int \mathcal{D}[\Delta_s,\Delta_s^*]\,
e^{-S^{(s)}_{\mathrm{HS}}},
\end{align}
with
\begin{align}
S^{(s)}_{\mathrm{HS}}
=
\int_0^\beta d\tau\Bigg[
\frac{|\Delta_s(\tau)|^2}{|U|}
-\frac{\Delta_s(\tau)}{\sqrt{N}}\sum_{\mathbf{k}}
\bar\psi_{\mathbf{k}\uparrow}(\tau)\bar\psi_{-\mathbf{k}\downarrow}(\tau)
-\frac{\Delta_s^*(\tau)}{\sqrt{N}}\sum_{\mathbf{k}}
\psi_{-\mathbf{k}\downarrow}(\tau)\psi_{\mathbf{k}\uparrow}(\tau)
\Bigg].
\end{align}

\paragraph*{Equal-spin triplet $p$ channels.}
Similarly, for each $(\eta,\sigma)$ with $\eta=x,y$ and $\sigma=\uparrow,\downarrow$, the projected NN Cooper interaction reads
\begin{align}
\exp\Bigg[
\frac{2|V|}{N}\int_0^\beta d\tau
\sum_{\mathbf{k},\mathbf{k}'}
\phi_{p\eta}(\mathbf{k})\phi_{p\eta}(\mathbf{k}')
\bar\psi_{\mathbf{k}\sigma}(\tau)\bar\psi_{-\mathbf{k}\sigma}(\tau)\,
\psi_{-\mathbf{k}'\sigma}(\tau)\psi_{\mathbf{k}'\sigma}(\tau)
\Bigg]
\propto
\int \mathcal{D}[\Delta_{p\eta,\sigma},\Delta^*_{p\eta,\sigma}]\,
e^{-S^{(p)}_{\mathrm{HS}}},
\end{align}
where
\begin{align}
S^{(p)}_{\mathrm{HS}}
=
\int_0^\beta d\tau\Bigg[
\frac{|\Delta_{p\eta,\sigma}(\tau)|^2}{2|V|}
-\frac{\Delta_{p\eta,\sigma}(\tau)}{\sqrt{N}}\sum_{\mathbf{k}}
\phi_{p\eta}(\mathbf{k})\,
\bar\psi_{\mathbf{k}\sigma}(\tau)\bar\psi_{-\mathbf{k}\sigma}(\tau)
-\frac{\Delta^*_{p\eta,\sigma}(\tau)}{\sqrt{N}}\sum_{\mathbf{k}}
\phi_{p\eta}(\mathbf{k})\,
\psi_{-\mathbf{k}\sigma}(\tau)\psi_{\mathbf{k}\sigma}(\tau)
\Bigg].
\end{align}

\paragraph*{Quadratic fermionic action.}
After decoupling the $s$ and all equal-spin $p$ channels, the partition function becomes
\begin{align}
Z
=
\int \mathcal{D}[\Delta_s,\Delta_s^*]
\prod_{\eta=x,y}\prod_{\sigma=\uparrow,\downarrow}
\mathcal{D}[\Delta_{p\eta,\sigma},\Delta^*_{p\eta,\sigma}]
\int \mathcal{D}[\bar\psi,\psi]\,
\exp\Big[-S_{\mathrm{quad}}[\bar\psi,\psi;\Delta]\Big],
\end{align}
with the explicit quadratic action
\begin{align}
S_{\mathrm{quad}}[\bar\psi,\psi;\Delta]
=
\int_0^\beta d\tau\Bigg\{&
\sum_{\mathbf{k},\sigma}
\bar\psi_{\mathbf{k}\sigma}(\tau)\big(\partial_\tau+\xi_{\mathbf{k}\sigma}\big)\psi_{\mathbf{k}\sigma}(\tau)
+\frac{|\Delta_s(\tau)|^2}{|U|}
+\sum_{\eta,\sigma}\frac{|\Delta_{p\eta,\sigma}(\tau)|^2}{2|V|}
\nonumber\\
&-\frac{1}{\sqrt{N}}\sum_{\mathbf{k}}
\Big[\Delta_s(\tau)\,\bar\psi_{\mathbf{k}\uparrow}(\tau)\bar\psi_{-\mathbf{k}\downarrow}(\tau)
+\Delta_s^*(\tau)\,\psi_{-\mathbf{k}\downarrow}(\tau)\psi_{\mathbf{k}\uparrow}(\tau)\Big]
\nonumber\\
&-\frac{1}{\sqrt{N}}\sum_{\eta,\sigma}\sum_{\mathbf{k}}
\phi_{p\eta}(\mathbf{k})
\Big[\Delta_{p\eta,\sigma}(\tau)\,\bar\psi_{\mathbf{k}\sigma}(\tau)\bar\psi_{-\mathbf{k}\sigma}(\tau)
+\Delta^*_{p\eta,\sigma}(\tau)\,\psi_{-\mathbf{k}\sigma}(\tau)\psi_{\mathbf{k}\sigma}(\tau)\Big]
\Bigg\}.
\label{eq:S_quad_explicit}
\end{align}
For the Gaussian (Landau) expansion discussed in the main text, it is sufficient to consider static, uniform HS fields
$\Delta_s(\tau)\equiv \Delta_s$ and $\Delta_{p\eta,\sigma}(\tau)\equiv \Delta_{p\eta,\sigma}$.

\subsection*{D. Integrating out fermions and Landau expansion via $G_0+\Sigma_\Delta$}

After the Hubbard--Stratonovich (HS) decoupling in the $s$-wave singlet channel
and the NN equal-spin $p$-wave channels, the fermionic action is quadratic in the
Grassmann fields. Restricting to uniform ($\mathbf{q}=0$) and static HS fields,
$\Delta_s(\tau)\!=\!\Delta_s$ and $\Delta_{p\eta,\sigma}(\tau)\!=\!\Delta_{p\eta,\sigma}$,
the partition function takes the form
\begin{equation}
Z=\int \mathcal{D}[\Delta,\Delta^*]\,
e^{-S_B[\Delta,\Delta^*]}\,
\int \mathcal{D}[\bar\psi,\psi]\,
e^{-S_F[\bar\psi,\psi;\Delta,\Delta^*]},
\end{equation}
with the bosonic (HS) weight
\begin{equation}
S_B[\Delta,\Delta^*]=\beta\left(
\frac{|\Delta_s|^2}{|U|}
+\sum_{\eta=x,y}\sum_{\sigma=\uparrow,\downarrow}
\frac{|\Delta_{p\eta,\sigma}|^2}{2|V|}
\right),
\label{eq:SB}
\end{equation}
and $\beta=1/T$.

\paragraph{Nambu representation and anomalous self-energy.}
It is convenient to write the quadratic fermionic action in Nambu space.
We define the four-component Nambu spinor
\begin{equation}
\Psi_{\mathbf{k},n}\equiv
\begin{pmatrix}
\psi_{\mathbf{k}\uparrow}(i\omega_n)\\
\psi_{\mathbf{k}\downarrow}(i\omega_n)\\
\bar\psi_{-\mathbf{k}\uparrow}(-i\omega_n)\\
\bar\psi_{-\mathbf{k}\downarrow}(-i\omega_n)
\end{pmatrix},\qquad
\omega_n=(2n+1)\pi T,
\end{equation}
so that
\begin{equation}
S_F=\frac{1}{2}\sum_{\mathbf{k},n}
\bar\Psi_{\mathbf{k},n}\,
\bigl[\mathcal{G}_0^{-1}(\mathbf{k},i\omega_n)-\Sigma_\Delta(\mathbf{k})\bigr]\,
\Psi_{\mathbf{k},n}.
\label{eq:SF_nambu}
\end{equation}
Here the normal-state (free) Nambu Green's function is block diagonal,
\begin{equation}
\mathcal{G}_0^{-1}(\mathbf{k},i\omega_n)=
\begin{pmatrix}
i\omega_n-\xi_{\mathbf{k}\uparrow} & 0 & 0 & 0\\
0 & i\omega_n-\xi_{\mathbf{k}\downarrow} & 0 & 0\\
0 & 0 & i\omega_n+\xi_{\mathbf{k}\uparrow} & 0\\
0 & 0 & 0 & i\omega_n+\xi_{\mathbf{k}\downarrow}
\end{pmatrix},
\label{eq:G0_nambu}
\end{equation}
with $\xi_{\mathbf{k}\sigma}$ given in the main text.

The HS pairing fields generate an off-diagonal (anomalous) self-energy
\begin{equation}
\Sigma_\Delta(\mathbf{k})=
\begin{pmatrix}
0 & \hat\Delta(\mathbf{k})\\
\hat\Delta^\dagger(\mathbf{k}) & 0
\end{pmatrix},
\label{eq:SigmaDelta_def}
\end{equation}
where the $2\times 2$ gap matrix in spin space is
\begin{equation}
\hat\Delta(\mathbf{k})=
\begin{pmatrix}
\Delta_{\uparrow\uparrow}(\mathbf{k}) & \Delta_s\\
-\Delta_s & \Delta_{\downarrow\downarrow}(\mathbf{k})
\end{pmatrix}.
\label{eq:gapmatrix}
\end{equation}
In our channel restriction,
\begin{equation}
\Delta_{\sigma\sigma}(\mathbf{k})=\sum_{\eta=x,y}\Delta_{p\eta,\sigma}\,\phi_{p\eta}(\mathbf{k}),
\qquad
\phi_{p\eta}(\mathbf{k})=\sin k_\eta,
\label{eq:triplet_gap}
\end{equation}
and the odd parity $\phi_{p\eta}(-\mathbf{k})=-\phi_{p\eta}(\mathbf{k})$ ensures the
antisymmetry of equal-spin pairing. (For on-site $s$ wave we have $\phi_s(\mathbf{k})\equiv 1$.)

\paragraph{Integrating out fermions.}
Using the standard Gaussian Grassmann integral
$\int\mathcal{D}[\bar\psi,\psi]\exp[-\bar\psi A\psi]=\det A$,
Eq.~\eqref{eq:SF_nambu} yields (up to an irrelevant constant from Nambu doubling)
\begin{equation}
\int \mathcal{D}[\bar\psi,\psi]\,
e^{-S_F}
\propto
\exp\!\left[-\frac{1}{2}\Tr\ln\bigl(\mathcal{G}_0^{-1}-\Sigma_\Delta\bigr)\right],
\end{equation}
where $\Tr$ denotes a trace over momentum, Matsubara frequency, and internal Nambu/spin indices.
Therefore the effective action for the HS fields is
\begin{equation}
S_{\mathrm{eff}}[\Delta,\Delta^*]
=
S_B[\Delta,\Delta^*]
-\frac{1}{2}\Tr\ln\bigl[\mathcal{G}_0^{-1}-\Sigma_\Delta\bigr].
\label{eq:Seff}
\end{equation}

\paragraph{Landau expansion to quadratic order.}
To obtain the quadratic Landau coefficients, we expand Eq.~\eqref{eq:Seff} around the normal state:
\begin{align}
-\Tr\ln(\mathcal{G}_0^{-1}-\Sigma_\Delta)
&=
-\Tr\ln\mathcal{G}_0^{-1}
-\Tr\ln\!\left[\mathbb{1}-\mathcal{G}_0\Sigma_\Delta\right]\nonumber\\
&=
-\Tr\ln\mathcal{G}_0^{-1}
+\sum_{m=1}^{\infty}\frac{1}{m}\Tr\!\left[(\mathcal{G}_0\Sigma_\Delta)^m\right].
\label{eq:logexpansion}
\end{align}
The linear term ($m=1$) vanishes because $\Sigma_\Delta$ is purely off-diagonal in Nambu space
and the normal state has no anomalous expectation value. Keeping only $m=2$ gives
\begin{equation}
S_{\mathrm{eff}}^{(2)}=
\beta\,a_s(T,\alpha)|\Delta_s|^2
+\beta\sum_{\eta,\sigma}a_{p\eta,\sigma}(T,\alpha)|\Delta_{p\eta,\sigma}|^2,
\label{eq:Seff2}
\end{equation}
with
\begin{align}
a_s(T,\alpha) &= \frac{1}{|U|}-\Pi_s(T,\alpha), \label{eq:as_def}\\
a_{p\eta,\sigma}(T,\alpha) &= \frac{1}{2|V|}-\Pi_{p\eta,\sigma}(T,\alpha). \label{eq:ap_def}
\end{align}
Evaluating the trace in Eq.~\eqref{eq:logexpansion} at $m=2$ leads to the usual Cooper-channel
bubbles (computed solely from the free Green's functions
$G_{0,\sigma}(\mathbf{k},i\omega_n)=(i\omega_n-\xi_{\mathbf{k}\sigma})^{-1}$):
\begin{align}
\Pi_s(T,\alpha)
&=
\frac{1}{\beta N}\sum_{\mathbf{k},n}
G_{0,\uparrow}(\mathbf{k},i\omega_n)\,
G_{0,\downarrow}(-\mathbf{k},-i\omega_n),
\label{eq:Pi_s_def}\\
\Pi_{p\eta,\sigma}(T,\alpha)
&=
\frac{1}{\beta N}\sum_{\mathbf{k},n}
\phi_{p\eta}^2(\mathbf{k})\,
G_{0,\sigma}(\mathbf{k},i\omega_n)\,
G_{0,\sigma}(-\mathbf{k},-i\omega_n).
\label{eq:Pi_p_def}
\end{align}

\subsection*{E. Explicit bubble expressions}

Here we perform the Matsubara-frequency summation for the normal-state Cooper bubbles
entering the quadratic (Gaussian) Landau coefficients derived in Sec.~S1D.
Using the free Green's function
$G_{0,\sigma}(\mathbf{k},i\omega_n)=(i\omega_n-\xi_{\mathbf{k}\sigma})^{-1}$
with $\omega_n=(2n+1)\pi T$, and the even-in-momentum dispersion
$\xi_{-\mathbf{k}\sigma}=\xi_{\mathbf{k}\sigma}$, we obtain the standard identity
\begin{align}
T\sum_{\omega_n} G_{0,\sigma}(\mathbf{k},i\omega_n)\,G_{0,\sigma'}(-\mathbf{k},-i\omega_n)
&=T\sum_{\omega_n}\frac{1}{(i\omega_n-\xi_{\mathbf{k}\sigma})(-i\omega_n-\xi_{\mathbf{k}\sigma'})}
\nonumber\\
&=\frac{1-n_F(\xi_{\mathbf{k}\sigma})-n_F(\xi_{\mathbf{k}\sigma'})}
{\xi_{\mathbf{k}\sigma}+\xi_{\mathbf{k}\sigma'}}
\nonumber\\
&=\frac{\tanh(\xi_{\mathbf{k}\sigma}/2T)+\tanh(\xi_{\mathbf{k}\sigma'}/2T)}
{2(\xi_{\mathbf{k}\sigma}+\xi_{\mathbf{k}\sigma'})},
\label{eq:SM_matsubara_bubble}
\end{align}
where $n_F(x)=1/(e^{x/T}+1)$ is the Fermi function.

Inserting Eq.~(\ref{eq:SM_matsubara_bubble}) into the channel-resolved bubbles gives the explicit
susceptibilities used in the main text.
For the on-site singlet $s$ channel, $\phi_s(\mathbf{k})=1$ and $(\sigma,\sigma')=(\uparrow,\downarrow)$,
\begin{equation}
\Pi_s(T,\alpha)
=\frac{1}{N}\sum_{\mathbf{k}}
\frac{\tanh(\xi_{\mathbf{k}\uparrow}/2T)+\tanh(\xi_{\mathbf{k}\downarrow}/2T)}
{2(\xi_{\mathbf{k}\uparrow}+\xi_{\mathbf{k}\downarrow})}.
\label{eq:SM_Pi_s}
\end{equation}
For the equal-spin nearest-neighbor $p_\eta$ channels with $\phi_{p_\eta}(\mathbf{k})=\sin k_\eta$ and $\sigma'=\sigma$,
we obtain
\begin{equation}
\Pi_{p\eta,\sigma}(T,\alpha)
=\frac{1}{N}\sum_{\mathbf{k}}
\sin^2 k_\eta\,
\frac{\tanh(\xi_{\mathbf{k}\sigma}/2T)}{2\xi_{\mathbf{k}\sigma}},
\qquad \eta=x,y,\ \sigma=\uparrow,\downarrow.
\label{eq:SM_Pi_p}
\end{equation}
Equations~(\ref{eq:SM_Pi_s})--(\ref{eq:SM_Pi_p}) reproduce the explicit bubbles quoted in the main text.

\section*{S2. Anisotropy dependence of effective $s$- and $p$-wave pair distributions}

\begin{figure*}[ht!]
    \centering
    \includegraphics[width=1\linewidth]{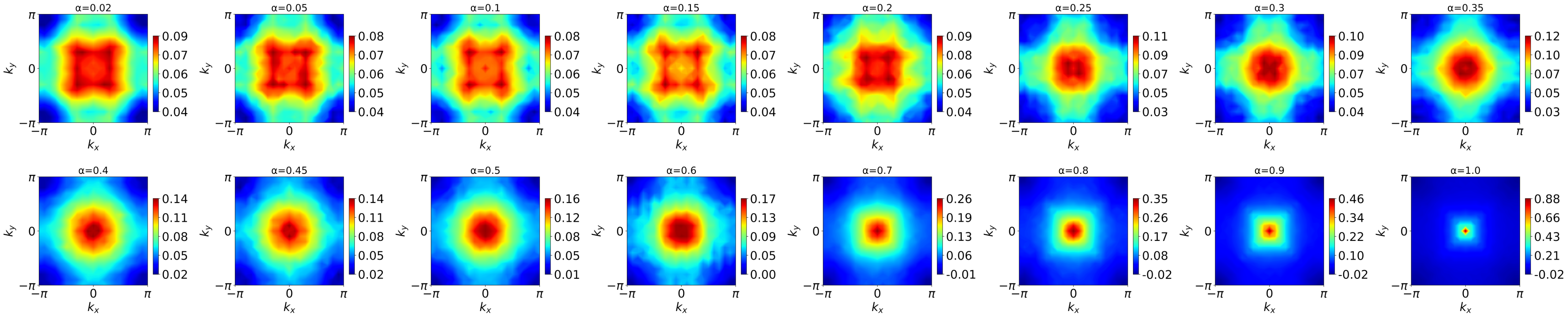}
   \caption{
(Color online) CPQMC results for the effective on-site $s$-wave pair momentum distribution
$N^{\mathrm{eff}}_{s\text{-pair}}(\mathbf{k})$ along the representative cut $V=-0.2$ at filling $n\simeq 0.85$
on a $20\times 20$ lattice, shown for anisotropies $0.02\le \alpha \le 1$.
The evolution from a broad Bose-surface feature at strong anisotropy (small $\alpha$) to a sharp peak at
$\Gamma$ as $\alpha\to 1$ visualizes the crossover from CPBM-like correlations to $s$-wave phase coherence.}
    \label{figS3}
\end{figure*}

\begin{figure*}[ht!]
    \centering
    \includegraphics[width=1\linewidth]{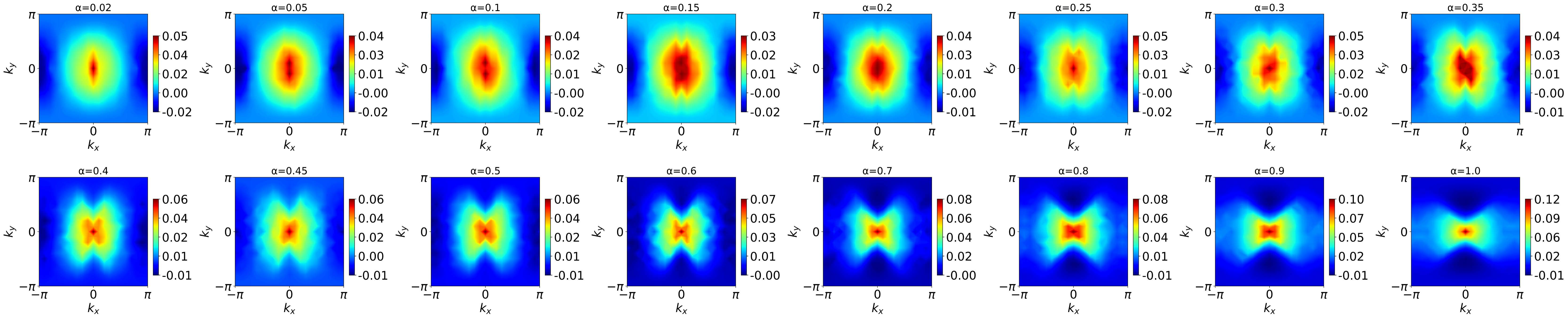}
    \caption{(Color online) Same as Fig.~S3, but for the $p$-wave vertex distribution
$N^{\mathrm{eff}}_{p\text{-pair}}(\mathbf{k})$ along the same cut $V=-0.2$.
While the overall magnitude changes only weakly with $\alpha$, strong anisotropy selects a preferred momentum-axis
orientation (polar axis) for the $p$-wave correlations, and the distribution gradually approaches a more
fourfold-symmetric pattern as $\alpha\to 1$.}
    \label{figS4}
\end{figure*}

In this section we examine how the momentum-resolved effective pair distributions
$N^{\mathrm{eff}}_{s\text{-pair}}(\mathbf{k})$ and $N^{\mathrm{eff}}_{p\text{-pair}}(\mathbf{k})$
respond to the hopping anisotropy $\alpha$.
All data are obtained from CPQMC simulations at filling $n\simeq 0.85$ on a $20\times 20$ lattice
for the representative cut $V=-0.2$.

Figure~S3 shows $N^{\mathrm{eff}}_{s\text{-pair}}(\mathbf{k})$ for a sequence of anisotropies
$0.02\le \alpha \le 1$.
In the strong-anisotropy limit (small $\alpha$), the distribution exhibits a broad, nearly square-shaped
Bose surface without a sharp condensation peak at $\Gamma$, consistent with the CPBM regime.
As $\alpha$ increases (i.e., the anisotropy is reduced), the Bose-surface feature gradually contracts
and the weight near $\Gamma$ is enhanced.
Approaching the isotropic limit $\alpha\to 1$, $N^{\mathrm{eff}}_{s\text{-pair}}(\mathbf{k})$
develops a pronounced peak at $\Gamma$, indicating condensation of Cooper pairs and the onset of
long-range $s$-wave phase coherence.
Conversely, reducing $\alpha$ suppresses the condensate peak and drives the system into the CPBM regime,
consistent with the evolution of the phase boundary in the main-text phase diagram.

Figure~S4 displays the corresponding $p$-wave vertex distribution $N^{\mathrm{eff}}_{p\text{-pair}}(\mathbf{k})$
along the same cut $V=-0.2$.
In contrast to the $s$-wave case, the overall magnitude of the $p$-wave signal varies only moderately with $\alpha$,
while its momentum-space structure is strongly reshaped.
For small $\alpha$, the distribution forms an elongated feature along a single momentum axis, reflecting that
spin-dependent hopping anisotropy lifts the near-degeneracy between the $p_x$ and $p_y$ channels and selects a
preferred orientation (a polar axis) for triplet correlations.
With increasing $\alpha$, this anisotropic feature broadens and evolves toward a more fourfold-symmetric pattern,
consistent with the reduced spin-dependent Fermi-surface deformation as $\alpha\to 1$.
Overall, these results support the main-text conclusion that hopping anisotropy primarily suppresses on-site
$s$-wave coherence and controls the orientation/morphology of $p$-wave correlations, rather than simply amplifying
their total weight.

 

\bibliography{ref}